\documentstyle[twocolumn,aps,epsf]{revtex}

\begin{document}
\draft
\preprint{HEP/123-qed}
\title{Probabilistic Quantum Memories}
\author{C. A. Trugenberger}
\address{InfoCodex,\\
chemin du Petit-Saconnex 28, CH-1209 Gen\`eve, Switzerland}
%\author{C. A. Trugenberger\cite{bylineb}}
%\author{pippo}
\address{e-mail: ca.trugenberger@bluewin.ch}
\date{\today}
\maketitle
\begin{abstract}
Typical address-oriented computer memories cannot recognize incomplete or noisy
information. Associative (content-addressable) memories solve this problem but
suffer from severe capacity shortages. I propose a model of a quantum memory
that solves both problems. The storage capacity is exponential in the number of
qbits and thus optimal. The retrieval mechanism for incomplete or noisy inputs
is probabilistic, with postselection of the measurement result. The output is
determined by a probability distribution on the memory which is peaked around
the stored patterns closest in Hamming distance to the input. 
\end{abstract}

\narrowtext
Quantum computation \cite{steane} is normally associated with new complexity
classes which are inaccessible (in polynomial time) to classical Turing
machines. In other words, quantum algorithms \cite{pittenger} can drastically
speed up the solution of tasks with respect to their classical counterparts, the
paramount examples being Shor's factoring algorithm \cite{shor} and Grover's
search algorithm \cite{grover}.

There is, however, another aspect of quantum computation which represents a big
improvement upon its classical counterpart. In traditional computers the storage
of information requires setting up a lookup table (RAM). The main disadvantage
of this address-oriented memory system lies in its rigidity. Retrieval of
information requires a precise knowledge of the memory address and, therefore,
incomplete or noisy inputs are not permitted.

In order to address this shortcoming, models of associative (or
content-addressable) memories \cite{neuralnetworks} were introduced. Here,
recall of information is possible on the basis of partial knowledge of their
content, without knowing the storage location. These are examples of collective
computation on neural networks \cite{neuralnetworks}, the best known example
being the Hopfield model \cite{hopfield} and its generalization to a
bidirectional associative memory \cite{kosko}. 

While these models solve the problem of recalling incomplete or noisy inputs,
they suffer from a severe capacity shortage. Due to the phenomenon of crosstalk,
which is essentially a manifestation of the spin glass transition \cite{parisi}
in the corresponding spin systems, the maximum number of binary patterns that
can be stored in a Hopfield network of n neurons is $p_{max} \simeq 0.14 \ n$
\cite{neuralnetworks} . While various possible improvements can be introduced
\cite{neuralnetworks}, the maximum number of patterns remains linear in the
number of neurons, $p_{max} = O(n)$.

In this paper I show that quantum mechanical entanglement provides a natural
mechanism for both improving dramatically the storage capacity of associative
memories and retrieving noisy or incomplete information. Indeed, the number of
binary patterns that can be stored in such a {\it quantum memory} is exponential
in the number $n$ of qbits, $p_{max} = 2^n$, i.e. it is optimal in the sense
that all binary patterns that can be formed with $n$ bits can be stored. The
retrieval mechanism is probabilistic, with postselection of the measurement
result. This means that one has to repeat the retrieval algorithm until 
a threshold $T$ is reached or the
measurement of a control qbit yields a given result. 
In the former case the input is not recognized. In the latter case, instead, 
the output is determined itself by a probability distribution 
on the memory which is peaked around the stored 
patterns closest (in Hamming distance) to the input.
The efficiency of this information retrieval mechanism depends on the
distribution of the stored patterns. Recognition efficiency is best when the
number of stored patterns is very large while identification efficiency
is best for isolated patterns which are very different from all other ones,
both very intuitive features.

Let me start by describing the elementary quantum gates \cite{pittenger}
that I will use in the rest of the paper. First of all there are the
single-qbit gates NOT, represented by the first Pauli matrix $\sigma_1$, and
H (Hadamard), with the matrix representation
\begin{equation}
H = {1\over \sqrt{2}} \ \left( \matrix{1 & 1\cr
1 & -1\cr} \right) \ .
\label{adda}
\end{equation}
Then, I will use extensively the two-qbit XOR (exclusive OR) gate, which
performs a NOT on the second qbit if and only if the first one is in state
$|1\rangle$. In matrix notation this gate is represented as
${\rm XOR} = {\rm diag} \left( 1, \sigma_1 \right)$, where $1$ denotes a
two-dimensional identity matrix and $\sigma _1$ acts on the components
$|01\rangle$ and $|11\rangle$ of the Hilbert space. The 2XOR, or Toffoli gate
\cite{gates} is the three qbit generalization of the XOR gate: it performs a
NOT on the third qbit if and only if the first two are both in state
$|1\rangle$. In matrix notation it is given by 
${\rm 2XOR} = {\rm diag} \left( 1, 1, \sigma_1 \right)$. In the storage
algorithm I shall make use also of the nXOR generalization of these
gates, in which there are n control qbits. This gate is also used in
the subroutines implementing the oracles underlying Grover's algorithm
\cite{pittenger} and can be realized using unitary maps affecting only few
qbits at a time \cite{gates}, which makes it feasible. 
All these are standard gates. In addition to them I introduce the two-qbit
controlled gates
\begin{eqnarray}
CS^i &&= |0\rangle \langle 0| \otimes 1 + |1\rangle \langle 1|
\otimes S^i \ ,
\nonumber \\
S^i &&= \left( \matrix{\sqrt{i-1\over i}&1\over \sqrt{i}\cr
-1\over{\sqrt{i}}&\sqrt{i-1\over i}\cr} \right) \ ,
\label{addb}
\end{eqnarray}
for $i=1, \dots, p$. These have the matrix notation $CS^i = {\rm diag}
\left( 1, S^i \right)$. For all these gates I shall indicate by subscripts
the qbits on which they are applied, the control qbits coming always first.  

Given $p$ binary patterns $p_i$ of length $n$, it is not difficult to imagine
how a quantum memory can store them. Indeed, such a memory is naturally 
provided by the following superposition of $n$ entangled qbits:
\begin{equation}
|M\rangle = {1\over \sqrt{p}}\ \sum_{i=1}^p \ |p^i\rangle \ .
\label{a}
\end{equation}
The only real question is how to generate this state unitarily from a 
simple initial state of $n$ qbits. 
To this end one can use the algorithm proposed in \cite{ventura} . 
Here, however, I shall propose a simplified version.

In constructing $|M\rangle$ I shall use three registers: a first register $p$ of
$n$ qbits in which I will subsequently feed the patterns $p^i$ to be stored, a
utility register $u$ of two qbits prepared in state $|01\rangle$, 
and another register $m$
of $n$ qbits to hold the memory. This latter will be initially prepared in state
$|0_1, \dots, 0_n\rangle$. The full initial quantum state is thus
\begin{equation}
|\psi _0^1\rangle = |p^1_1, \dots p^1_n; 01; 0_1, \dots, 0_n\rangle \ .
\label{c}
\end{equation}
The idea of the storage algorithm is to separate this state 
into two terms, one corresponding to the already stored patterns, and another
ready to process a new pattern. These two parts will be distinguished by the
state of the second utility qbit $u_2$: $|0\rangle$ for the stored patterns 
and $|1\rangle$ for the processing term. 

For each pattern $p^i$ to be stored 
one has to perform the operations described below:
\begin{equation}
|\psi _1^i \rangle = \prod_{j=1}^n 
\ 2XOR_{p_j^i u_2 m_j} \ |\psi _0^i\rangle \ .
\label{addc}
\end{equation}
This simply copies pattern $p^i$ into the memory register of the processing
term, identified by $|u_2\rangle = |1\rangle$.
\begin{eqnarray}
|\psi _2^i \rangle &&= \prod_{j=1}^n 
\ NOT_{m_j} \ XOR_{p_j^i m_j} \ |\psi _1^i\rangle \ ,
\nonumber \\
|\psi _3^i \rangle &&= nXOR_{m_1 \dots m_n u_1} |\psi_2^i \rangle \ .
\label{addd}
\end{eqnarray}
The first of these operations makes all qbits of the memory register
$|1\rangle$'s when the contents of the pattern and memory registers are
identical, which is exactly the case only for the processing term. Together,
these two operations change the first utility qbit $u_1$ of the processing
term to a $|1\rangle$, leaving it unchanged for the stored patterns term.
\begin{equation}
|\psi_4^i\rangle = CS^{p+1-i}_{u_1 u_2} \ |\psi _3^i\rangle \ .
\label{adde}
\end{equation}
This is the central operation of the storing algorithm. It separates out the
new pattern to be stored, already with the correct normalization factor.
\begin{eqnarray}
|\psi _5^i \rangle &&= nXOR_{m_1 \dots m_n u_1} |\psi_4^i \rangle \ ,
\nonumber \\
|\psi _6^i \rangle &&= \prod_{j=n}^1
\ XOR_{p_j^i m_j} \ NOT_{m_j}\  |\psi _5^i\rangle \ .
\label{addf}
\end{eqnarray}
These two operations are the inverse of eqs.(\ref{addd}) and restore the
utility qbit $u_1$ and the memory register $m$ to their original values. 
After these operations on has
\begin{equation}
|\psi _6^i \rangle = 
{1\over \sqrt{p}}\ \sum_{k=1}^i |p^i;00;p^k\rangle + \sqrt{p-i\over p}
|p^i;01;p^i\rangle \ .
\label{d}
\end{equation}
With the last operation,
\begin{equation}
|\psi_7^i\rangle = \prod_{j=n}^1\ 2XOR_{p^i_j u_2 m_j}
\ |\psi_6^i\rangle \ ,
\label{addg}
\end{equation}
one restores the third register $m$ of the processing term, the second term
in eq.(\ref{d}) above, to its initial value $|0_1, \dots, 0_n\rangle$. At this
point one can load a new pattern into register $p$ and go through the 
same routine as just described. At the end of the whole process, 
the $m$-register is exactly in state $|M\rangle$, eq. (\ref{a}).

Assume now one is given a binary input $i$, which might be, e.g. a corrupted
version of one of the patterns stored in the memory. The first step of the 
information recall process is to make a copy of the memory $|M\rangle$ to be
used in the retrieval algorithm described below. Due to the no-cloning theorem
\cite{zurek},  this cannot be done deterministically (i.e. using only unitary
operations); a faithful copy of $|M\rangle$ can be obtained only with 
a probabilistic cloning machine \cite{cloning}.
I shall thus assume the availability of a probabilistic cloning machine
for which $|M\rangle$ is one of the set of linearly independent states that
can be copied.

The retrieval algorithm requires also three registers. The first register
$i$ of n qbits contains the input pattern; 
the second register $m$, also of n qbits,
contains the memory $|M\rangle$; finally there is a single qbit control
register $c$ initialized to the state $\left( |0\rangle + |1\rangle \right)/
\sqrt{2}$. The full initial quantum state is thus
\begin{eqnarray}
|\psi _0\rangle &&= {1\over \sqrt{2p}} \ \sum_{k=1}^p |i_1, \dots, i_n; 
p^k_1, \dots, p^k_n; 0\rangle 
\nonumber \\
&&+{1\over \sqrt{2p}} \ \sum_{k=1}^p |i_1, \dots, i_n; 
p^k_1, \dots, p^k_n; 1\rangle \ .
\label{e}
\end{eqnarray}
I now apply to it the following combination of quantum gates:
\begin{equation}
|\psi _1\rangle = \prod_{k=1}^n \ NOT_{m_k} 
\ XOR_{i_k m_k} |\psi _0\rangle \ ,
\label{f}
\end{equation}
where, as before, the subscripts on the
gates refer to the qbits on which they are applied. As a result of
this, the memory register qbits
are in state $|1\rangle$ if $i_j$ and $p^k_j$ are identical
and $|0\rangle$ otherwise:
\begin{eqnarray}
|\psi _1\rangle &&= {1\over \sqrt{2p}} \ \sum_{k=1}^p |i_1, \dots, i_n; 
d^k_1, \dots, d^k_n; 0\rangle 
\nonumber \\
&&+{1\over \sqrt{2p}} \ \sum_{k=1}^p |i_1, \dots, i_n; 
d^k_1, \dots, d^k_n; 1\rangle \ ,
\label{g}
\end{eqnarray}
where $d^k_j = 1$ if and only if  $i_j=p^k_j$ and $d^k_j=0$ otherwise.

Consider now the following Hamiltonian:
\begin{eqnarray}
{\cal H} &&= \left( d_H \right)_m \otimes \left( \sigma_3 \right)_c \ ,
\nonumber \\
\left( d_H \right)_m && = \sum_{k=1}^n 
\left( {\sigma_3 + 1\over 2} \right) _{m_k}\ ,
\label{h}
\end{eqnarray}
where $\sigma _3$ is the third Pauli matrix.
${\cal H}$ measures the number of 0's in register $m$, with a plus sign if $c$
is in state $|0\rangle$ and a minus sign if $c$ is in state $|1\rangle$. Given
how I have prepared the state $|\psi _1\rangle$, this is nothing else than the
number of qbits which are different in the input and memory registers $i$ and
$m$. This quantity is called the {\it Hamming distance} and represents the
(squared) Euclidean distance between two binary patterns. 

Every term in the superposition (\ref{g}) is an eigenstate of ${\cal H}$ with a
different eigenvalue. Applying thus the unitary operator ${\rm exp} 
(i \pi {\cal H}/2n)$ to $|\psi _1\rangle$ one obtains
\begin{eqnarray}
|\psi _2\rangle &&= {\rm e}^{i{\pi \over 2n}{\cal H}} \ |\psi_1\rangle \ ,
\label{i} \\
|\psi_2\rangle &&= {1\over \sqrt{2p}} \sum_{k=1}^p {\rm e}^{i{\pi\over 2n}
d_H\left( i, p^k\right)}
|i_1, \dots, i_n; d^k_1, \dots, d^k_n; 0\rangle 
\nonumber \\
&&+ {1\over \sqrt{2p}} \sum_{k=1}^p {\rm e}^{-i{\pi\over 2n}
d_H\left( i, p^k\right)}
|i_1, \dots, i_n; d^k_1, \dots, d^k_n; 1\rangle \ ,
\nonumber
\end{eqnarray}
where $d_H\left( i, p^k \right)$ denotes the Hamming distance bewteen the input
$i$ and the stored pattern $p^k$. 

In the final step I restore the memory gate to the state 
$|M\rangle$ by applying
the inverse transformation to eq. (\ref{f}) and I 
apply the Hadamard gate (\ref{adda}) 
to the control qbit, thereby obtaining
\begin{eqnarray}
|\psi _3\rangle &&= H_c \prod_{k=n}^1 XOR_{i_k m_k}  
\ NOT_{m_k} \ |\psi_2\rangle \ ,
\label{l} \\
|\psi_3\rangle &&= {1\over \sqrt{p}} \sum_{k=1}^p {\rm cos}\  {\pi \over
2n} d_H\left( i, p^k\right) 
|i_1, \dots, i_n; p^k_1, \dots, p^k_n; 0\rangle 
\nonumber \\
&&+ {1\over \sqrt{p}} \sum_{k=1}^p {\rm sin}\  {\pi \over 2n}
d_H\left( i, p^k\right) 
|i_1, \dots, i_n; p^k_1, \dots, p^k_n; 1\rangle .
\nonumber
\end{eqnarray}

This concludes the deterministic part of the information retrieval process.
At this point one needs a measurement 
of the control qbit $c$. The probabilities
for this to be in states $|0\rangle$ and $|1\rangle$  are given by the
expressions
\begin{eqnarray}
P(|c\rangle = |0\rangle ) &&= \sum_{k=1}^p \ {1\over p}\ 
{\rm cos}^2 \left( {\pi \over 2n} d_H\left( i, p^k\right) \right) \ ,
\label{ma} \\
P(|c\rangle = |1\rangle ) &&= \sum_{k=1}^p \ {1\over p}\ 
{\rm sin}^2 \left( {\pi \over 2n} d_H\left( i, p^k\right) \right) \ .
\label{mb}
\end{eqnarray}
If the input pattern is very different from all stored patterns, one has a high
probability of measuring $|c\rangle = |1\rangle$. On the contrary, an input 
pattern close to all stored patterns leads to a high probability of measuring
$|c\rangle = |0\rangle$. One can thus set a threshold $T$: if $T$ repetitions
of the retrieval algorithm all lead to a measurement $|c\rangle = |1\rangle$
one classifies the input $i$ as {\it non-recognized}. If one gets a measurement
$|c\rangle = |0\rangle$ before the threshold is reached, instead, one classifies
the input $i$ as {\it recognized} and one can proceed to a measurement of the
memory register to identify it. This measurement yields pattern $p^k$ with
probability
\begin{equation}
P\left( p^k \right) = {1\over p P(|c\rangle = |0\rangle)} \ {\rm cos}^2
\left( {\pi \over 2n} d_H \left( i, p^k \right) \right) \ .
\label{n}
\end{equation}
This probability is peaked around those patterns which have the smallest
Hamming distance to the input. The highest probability of retrieval is thus
realized for that (those) pattern which is most similar to the input.

What about the efficiency of this information retrieval mechanism? Contrary
to any classical counterpart, this efficiency depends here on two features:
the threshold $T$ determining {\it recognition} and the shape of the probability
distribution in eq.(\ref{n}), determining the {\it identification}. The
threshold $T$ should be optimally chosen according to the probabilities in
eqs.(\ref{ma},\ref{mb}) and depends thus 
on the distribution of the stored patterns.
Indeed, the probability of recognition is 
determined by comparing (squared) cosines and
sines of the distances to the stored patterns. It is thus clear that the worst
case for recognition is the situation in which there is an isolated pattern,
with the remaining patterns forming a tight cluster spanning all the largest
distances to the first one. Let me suppose that $p=O\left( n^x\right)$,
$x\ll n$, and assume for simplicity that $p=1+\sum_{k=0}^x {n\choose k}$ and the
distribution is such that exactly all patterns of distances $d_H = n, n-1,
\dots, n-x$ to one isolated pattern are stored. If one presents exactly this
isolated pattern as input, one of the (squared) cosines in eq.(\ref{ma}) is 1,
while the rest all take the smallest possible values, giving
\begin{equation}
P(|c\rangle = |0\rangle) > {1\over p} + {\pi^2 \over 4n^2} \ .
\label{o}
\end{equation}
In order to have the best recognition efficiency also in this worst case, one
should therefore choose the threshold $T=O(n)$ for $x=1$ and $T=O\left( n^2
\right)$ for $n\gg x\ge 2$. While this entails a large number of repetitions,
it is still polynomial in the number $n$ of qbits and thus tractable. Note
also that the required threshold diminishes when the number of stored patterns
becomes very large, since, in this case, the distribution of patterns becomes
necessarily more homogeneous. Indeed, for the maximal number of stored patterns
$p=2^n$ one has $P(|c\rangle = |0\rangle) = 1/2$ and the recognition efficiency
becomes also maximal, as it should be. In the general case one can initially
estimate the $p$ recognition probabilities of the patterns by setting 
$i=p^k$ for $k=1,\dots , p$ in eq.(\ref{ma}). Letting $P_{\rm min}$ be the
smallest of these, one can once and for all choose the threshold $T$ of this
memory as the nearest integer to $1/P_{\rm min}$. I do not discuss here a
possible quantum speed-up of this calculation since the main point of the
present paper is the exponential storage capacity with retrieval of noisy
inputs.

While the recognition efficiency depends on comparing 
(squared) cosines and sines of the same distances in the distribution, the
identification efficiency of eq.(\ref{n}) depends on comparing the (squared)
cosines of the different distances in the distribution.
Specifically, it is best when one of the distances is zero, while all others are
as large as possible, such that the probability of retrieval is completely
peaked on one pattern. As a consequence, the
identification efficiency is best when the recognition efficiency is worst 
and viceversa.  

Having described at length the information retrieval mechanism for complete,
but possibly corrupted patterns, it is easy to incorporate also
incomplete ones. To this end assume that only $q<n$ qbits of the input are 
known and let me denote these by the indices $\{k1, \dots, kq\}$. After
assigning the remaining qbits randomly, there are two possibilities. One can
just treat the resulting complete input as a noisy one and proceed as above or,
better, one can limit the operator $\left( d_H \right)_m$ in the 
Hamiltonian (\ref{h}) to
\begin{equation}
\left( d_H \right)_m = \sum_{i=1}^q \ \left( {\sigma_3+1\over 2} 
\right)_{m_{ki}} \ ,
\label{p}
\end{equation}
so that the Hamming distances to the stored patterns are computed on the
basis of the known qbits only. After this the pattern recall process continues
exactly as described above. This second possibility has the advantage that it
does not introduce random noise in the similarity measure but it has the
disadvantage that the operations of the memory have to be adjusted to the
inputs.

This brings me to the last point, the feasibility of the described algorithms.
In this context I would like to point out that, in addition to the standard
NOT, H (Hadamard), XOR, 2XOR (Toffoli) and nXOR gates
\cite{pittenger} I have introduced 
only the two-qbit gates $CS^i$ in eq.
(\ref{addb}) and the unitary 
operator ${\rm exp}\left( i\pi {\cal H}/2n \right)$.
It remains thus only to show that this latter can be realized by simple gates
involving few qbits. To this end I introduce the single-qbit gate
\begin{equation}
U = \left( \matrix{{\rm e}^{i{\pi\over 2n}}&0\cr
0&1\cr} \right) \ ,
\label{q}
\end{equation}
and the two-qbit controlled \cite{pittenger} gate
\begin{equation}
CU^{-2} = |0\rangle \langle 0| \otimes 1 + |1\rangle \langle 1|
\otimes U^{-2} \ .
\label{r}
\end{equation}
It is then easy to check that ${\rm exp}\left( i\pi {\cal H}/2n \right)$ can
be realized as follows:
\begin{equation}
{\rm e}^{i{\pi \over 2n} {\cal H}} \ |\psi _1\rangle = 
\prod_{i=1}^n \left( CU^{-2} \right) _{c m_i} \ \prod_{j=1}^n U_{m_j}
\ |\psi_1\rangle \ ,
\label{t}
\end{equation}
where $c$ is the control qbit in the first series of gates. Essentially, this
means that one implements first ${\rm exp}\left( i\pi  d_H /2n \right)$
and then one
corrects by implementing ${\rm exp}\left( -i\pi  d_H /n \right)$ on that part of
the quantum state for which the control qbit $|c\rangle$ is in state
$|1\rangle$. This completes the proof of feasibility. 

It remains to point out that the information retrieval algorithm can be, in
principle, generalized by substituting the Hamiltonian (\ref{h}) with
\begin{equation}
{\cal H} = \left( f\left( d_H \right)
\right)_m \otimes \left( \sigma_3 \right)_c \ ,
\label{u}
\end{equation}
where $f$ is any function satisfying $f(0)=0$ and $f(n)=n$. Such a
generalization would above all have
an influence on the identification efficiency by changing the shape of the
probability distribution on the memory, which can be made narrower around the
input. One can also give different weights to different qbits by introducing a
non-trivial metric. The only restriction
on all these generalizations is, as always, the feasibility of the resulting
unitary evolution.

%\begin{figure}
%\centerline{\epsfysize=5.5cm\epsfbox{quartica.eps}}
%\caption{The roton-like minimum in the spectrum of transverse fluctuations
%for $I\gg R$.}
%\end{figure}

\end{document}